# Generation of "perfect" vortex of variable size and its effect in angular spectrum of the down-converted photons

M. V. Jabir[1,*], N. Apurv Chaitanya[1,2], A. Aadhi[1], G. K. Samanta[1]

[1] *Photonic Sciences Lab., Physical Research Laboratory, Navarangpura, Ahmedabad 380009, Gujarat, India*

[2] *Indian Institute of Technology-Gandhinagar, Ahmedabad 382424, Gujarat, India*

*\*Corresponding author: jabir@prl.res.in*

**Abstract:** *The "perfect" vortex is a new class of optical vortex beam having ring radius independent of its topological charge (order). One of the simplest techniques to generate such beams is the Fourier transformation of the Bessel-Gauss beam. The variation in ring radius of such vortices require Fourier lenses of different focal lengths and or complicated imaging setup. Here we report a novel experimental scheme to generate perfect vortex of any ring radius using a convex lens and an axicon. As a proof of principle, using a lens of focal length ƒ=200mm, we have varied the radius of the vortex beam across 0.3-1.18mm simply by adjusting the separation between the lens and axicon. This is also a simple scheme to measure the apex angle of an axicon with ease. Using such vortices we have studied non-collinear interaction of photons having orbital angular momentum (OAM) in spontaneous parametric down-conversion (SPDC) process and observed that the angular spectrum of the SPDC photons are independent of OAM of the pump photons rather depends on spatial profile of the pump beam. In the presence of spatial walk-off effect in nonlinear crystals, the SPDC photons have asymmetric angular spectrum with reducing asymmetry at increasing vortex radius.*

Spontaneous parametric down-conversion (SPDC)[1,2], one of the most important nonlinear processes, is of paramount interest especially in the field of quantum optics[3] for its intrinsic capability in generating entangled photon pairs. In this process, a photon (pump) of higher energy while interacting with second order nonlinear crystals splits into a pair of low energy photons (signal-idler) subject to energy conservation and phase matching. Depending upon the phase matching conditions inside the nonlinear crystals, the down converted photons are known to be emitted in variety of spatial distributions. For example, pump beam of Gaussian mode under type-I (where the signal and idler have same polarization but orthogonal to the pump polarization), non-collinear phase-matching produces down converted photons in an annular ring, with photon of each pair at diametrically opposite points. As such, these photon pairs are entangled in the spatial degrees of freedom[4] and other degrees of freedom including orbital angular momentum (OAM)[5]. The angular spectrum and hence the entanglement properties of these paired photons are highly influenced by different crystal parameters including birefringence and length, and the spatial structure of the pump beam[6].

Therefore, it is imperative to study the angular spectrum of the SPDC photons for different crystal parameters and pump beams of different spatial structures and OAM.

Optical vortices having phase singularities (phase dislocations) in the wavefront, carries vanishing intensity at the singular point. Due to the screw-like (helical) phase structure around the point of singularity, such beams carries OAM. The characteristic phase distribution of optical vortices is represented as *exp(ilθ)*, where θ is the azimuthal angle. The integer *l* is called the topological charge or the order of the vortex. It has been observed[7] that the OAM of the classical beam can be transferred to the heralded single photon generated through SPDC process. Therefore, it is important to study the spatial distribution of the SPDC photons while pumped with different order optical vortices. Recently we have reported[8] that unlike Gaussian beam, the vortex pump beams produces SPDC photons in asymmetric annular rings and the asymmetry increases with the order of the vortex. As the divergence[9] and the radius of the vortex beam increases with its order[10], it is difficult to ascertain the contribution of the spatial distribution of the vortex beam and it's OAM in the asymmetry of SPDC ring separately.

Recent development in the field of structured beams have resulted in a special class of optical vortices, known as perfect vortices[11], where the radius of the vortex ring is independent of its order. The experimentally realizable annular field profile of the perfect vortex with topological charge *l* at a fixed propagation distance may be written as[12]

$$E(\rho, \theta) \equiv exp\left(-\frac{(\rho-\rho_r)^2}{\Delta\rho^2}\right) exp(il\theta) \qquad (1)$$

where, $(\rho, \theta)$ are polar coordinates, $\rho_r$ and $\Delta\rho$ are the radius and annular width of the perfect vortex. Among different techniques[12,13,14], Fourier transformation of the Bessel-Gauss (BG) beam of different orders [14] is the simplest technique to generate perfect vortices. The axicon, an optical element with a conical surface, converts Laguerre-Gaussian (LG) beams into BG beams[15,16]. The order of the BG is the same as that of the input LG beams. The Gaussian (zero order LG beam) input beam results in $0^{th}$ order BG beam. Fourier transformation of such BG beams using a lens of focal length, *f*, results perfect vortex at the back focal plane with field amplitude distribution given as[13]

$$E(\rho, \theta) = i^{l-1} \frac{w_g}{w_o} exp(il\theta) exp\left(-\frac{\rho^2+\rho_r^2}{w_o^2}\right) I_l\left(\frac{2\rho_r\rho}{w_o^2}\right) \qquad (2)$$

where, $w_g$ is the beam waist radius of the Gaussian beam confining the BG beam, $I_l$ is the $l^{th}$ order modified Bessel function of first kind, and $2w_o$ ($w_o = 2f/kw_g$, the Gaussian beam waist at the focus) is the annular width of the perfect vortex of ring radius governed by the relation[13],

$$\rho_r = f sin((n-1)\alpha) \qquad (3)$$

where, *n* and *α* are the refractive index and base angle of the axicon respectively. As evident from Eqn. (3), for given axicon parameters, the radius of the perfect vortex is constant for a lens and linearly varying with the focal length of the Fourier transforming lens.

However, for all practical purposes[12,17,18], one need to vary the radius of the vortex ring, which requires either a number of Fourier transforming lenses of different focal lengths and or imaging systems of different demagnification factors. As such, both of these options can provide variation (only discrete values) in the vortex ring radius, definitely, at the cost of increased experimental complexity. Here we present a novel technique to generate perfect vortices with continuously varying annular ring radius. As a proof of principle, we have generated perfect vortices of topological charge as high as *l*=6 with annular ring radius continuously varying from 0.3-1.18 mm. This novel experimental scheme can potentially be used as a quick and simple technique to measure the apex angle of an axicon. Pumping a nonlinear crystal using such vortices of different ring radius we have studied the angular spectrum of the down converted photons and experimentally observed that the angular spectrum of the down converted photons is dictated by the spatial profile of the pump but not it's OAM. However, the asymmetry in the angular spectrum which is detrimental to the entanglement quality of the SPDC source[19], depends on the size of the input beam. Using the variable size perfect vortex we have experimentally found for given a nonlinear crystal (birefringence parameter and length) the input beam radius generating the symmetric angular spectrum of the down converted photons.

**Results**

**Theory**

According to the Fourier transformation theory[20], any object can be Fourier transformed by placing the object behind the lens at any arbitrary distance $D$ from the focal plane. The basic principle of the technique is pictorially represented in Fig. 1(b), where the object (axicon) placed at a distance ($f$ - $D$) behind the lens of focal length, $f$, is Fourier transformed into perfect vortex at the back focal plane for input vortex beams. The amplitude distribution of the perfect vortex at the back focal plane is given by

$$E(\rho,\theta) = i^{l-1} exp\left(i\frac{k\rho^2 f^2}{2D^3}\right) \frac{w_g}{w_D} exp(il\theta) exp\left(-\frac{\left[\rho\frac{f}{D}\right]^2 + \rho_r^2}{w_D^2}\right) I_l\left(\frac{2\rho_r\left[\rho\frac{f}{D}\right]}{w_D^2}\right) \quad (4)$$

In the current system configuration, the width of the perfect vortex is $2w_D = 1.65 \times 2w_o$ [14]. For large value of $\rho_r$, the modified Bessel function, $I_l$ can be approximated to[13]

$$I_l\left(\frac{2\rho_r\left[\rho\frac{f}{D}\right]}{w_D^2}\right) \sim exp\left(\frac{2\rho_r\left[\rho\frac{f}{D}\right]}{w_D^2}\right) \quad (5)$$

Therefore, Eqn. (4) can be reduced to

$$E(\rho,\theta) = i^{l-1} exp\left(i\frac{k\rho^2 f^2}{2D^3}\right) \frac{w_g}{w_D} exp\left(-\frac{\left(\rho\frac{f}{D} - \rho_r\right)^2}{w_D^2}\right) exp(il\theta) \quad (6)$$

The Eqn. (6) has similar functional form as Eqn. (1) with additional quadratic phase factor and a scale factor. As we are dealing with the intensity of the perfect vortex beam at focal plane, the quadratic phase does not make any difference in our current study. The perfect vortex ring is formed at, $\rho = \frac{D}{f}\rho_r$ with a modified ring radius,

$$\rho_r = D\,sin((n-1)\alpha) \quad (7)$$

As evident from Eqn. (7), the radius of the perfect vortex can be varied simply by moving the axicon away from the back focal plane of the lens. For ring radius $\rho_r$ to be zero the axicon required to be placed at the focal plane of the lens which is impractical for many applications. On the other hand, for $D\rightarrow 0$, the Fourier transform relation breaks down. Therefore, the value of $D$ can be varied practically within the range of $0<D<f$.

**Experiment**

The schematic of the experimental setup is shown in Fig. 1 (a). A UV diode laser (Toptica- TopMode 405) of 100 mW power at 405nm with spectral width ~12 MHz is used as the pump source. Two spiral phase plates, SPP1 and SPP2 are used to convert Gaussian beam into optical vortices (LG beams) of order, $l$=1 and $l=2$, respectively, at conversion efficiency >95%. Due to limited availability of SPPs producing vortices of order $l>2$, using a vortex-doubler setup[21] (see methods) we have generated optical vortices of orders (topological charges) up to $l=6$. The axicon with apex angle $178^o$ converts the vortex beam of order $l$ into BG beam of the same order. The Fourier transforming lens ($f=200$ mm) placed before the axicon with a separation of $f-D$ ($D$ is the distance of the axicon from Fourier plane) produces the perfect vortices at its back focal plane. The 5 mm long $\beta$-BaB$_2$O$_4$ (BBO) nonlinear crystal converts the pump photons at 405 nm into degenerate down converted photons at 810 nm in non-collinear phase-matching at $3^o$ with respect to the pump beam direction. Due to circular symmetry, the photons are generated in a cone of apex angle $6^o$ and the photons of each pair are situated in diametrically opposite points. A second lens, $f=50$ mm Fourier transforms the generated SPDC ring which is recorded using EMCCD after filtering the pump photon using an interference filter of 10 nm bandwidth centred at 810 nm.

**Discussions**

To verify the generation of perfect vortex, we have recorded the intensity profile of the beam at the back focal plane of the Fourier transforming lens ($f=200$ mm) using a CCD camera with the results shown in Fig. 2. The axicon is placed at D=160 mm before the Fourier plane. As evident from the intensity profile of the vortex beams of topological charge $l=1$ and $l=5$ shown in Fig. 2 (a) and 2 (b) respectively, both the beams have same annular ring radius, $\rho_r=1.18$ mm confirming the generation of vortices having annular ring radius independent of topological charge, perfect vortex. To confirm the vorticity and order of the perfect vortex, the vortex beam is interfered with a Gaussian beam at 405 nm with interference pattern shown in second column of Fig. 2. A close observation of the interference pattern of Figs. 2(c) and 2(d) reveals the characteristic spiral patterns at the annular ring of the beams corresponding to vortex orders $l=1$ and $l=5$ respectively. Unlike the spiral fringe pattern as observed in the interference of normal vortex beam with Gaussian beam, the spiral fringes in Figs. 2(c) and 2(d) is not visible at the center of the interference pattern due to the large dark core of the perfect vortices. Such observation clearly ascertain the capability of our experimental scheme in generating high-quality perfect optical vortices even at higher values of topological charges. Using Eqn. 6 along with the experimental parameters we have theoretically calculated the intensity profile of the perfect vortex and their interference pattern with Gaussian beam with the results shown in third (e-f) and fourth (g-h) column of Fig. 2 respectively. We can clearly see a close matching between theoretical and experimental results.

With the successful generation of perfect vortices with orders as high as $l=6$, we experimentally confirmed the variation of perfect vortex radius on demand. As a proof of principle, we recorded the intensity pattern of the perfect vortex of topological charge $l=3$ at the Fourier plane for three different axicon positions, D=40, 100 and 160 mm, with the results shown in Fig. 3. As evident from first row (a)-(c) of Fig. 3, the annular ring has radius of =0.3 mm, 0.75 mm and 1.18 mm for the $D$- value of 40 mm, 100 mm and 160 mm respectively. In principle the annular ring radius can further be reduced by decreasing the D-value, however, due to the mechanical constrain arising from the CCD camera structure and related optical components (attenuators) we could not record experimentally the perfect vortex of ring radius, $\rho_r<0.3$ mm by placing the axicon with separation D<40mm. To avoid such mechanical constrain in recording the vortex ring radius for D<40 mm, which is beyond the scope of present study, one can easily image the Fourier transforming plane to other measurement planes. On the other hand, further increase in vortex ring radius requires Fourier transforming lenses of longer focal lengths. Using Eq. (6) along with the experimental parameters, we have also theoretically calculated the intensity profile of the vortex of $l=3$ as shown in second row (d)-(f) of Fig. 3, in good agreement with the experimental results. However, we can observe a slight mismatch in the theoretical and experimental values of vortex ring radius which can be attributed to the inaccuracy of the apex angle used in theoretical study. To get more insight about the possibility of varying the ring radius of vortices of all orders, we have measured the ring radius of the vortices of three different orders $l=0$, 2 and 6 ($l=0$ is the Gaussian beam) with the results shown in Fig. 4. As evident from Fig. 4, the annular ring of the Gaussian as well the vortices of both orders have same radius for a fixed position of the axicon and increases linearly with the increase of D-value. Using Eq. (7) and axicon parameters (base angle $\alpha=1°$ and refractive index, n=1.5302 at 405 nm) we found a clear mismatch between theoretical (solid line) and experimental results. The theoretical and experimental variation of ring radius with axicon distance have slope of 9.23 x$10^{-3}$ and 7.33 x$10^{-3}$ respectively. From the experimentally measured slope (7.33 x$10^{-3}$) we found the apex angle of the axicon to be $178.4^0$ slightly higher than the value 178° presented by the manufacturer. To verify the manufacturing inaccuracy in the apex angle of the axicon, we have repeated the same experiment for three different laser wavelengths (532nm, 632 nm and 1064 nm). As expected, the annular ring radius increases with axicon position (D) away from the Fourier plane for all three wavelength at a slope of ~6.83 x$10^{-3}$, 7.05 x$10^{-3}$ and 7.05 x$10^{-3}$ for 532 nm, 632 nm and 1064 nm respectively. Using the refractive indices of the BK-7 glass we found the axicon to have apex angle of 178.4°±0.03° similar to the value measured previously. Such study also proves that the present experimental scheme (Fig. 1(b)) can be used as a quick and simple way of measuring the apex angle of an axicon. However, the measurement on the shape of the axicon tip and other axicon parameters required sophisticated equipment including optical profilometer[22].

To study of the contribution of OAM in the angular spectrum of the down converted photons, we pumped the nonlinear crystal (BBO) placed at the Fourier plane after the axicon using perfect vortices of different orders and ring radii. The angular spectrum or transverse momentum distribution of the down converted photons, obtained from the Fourier transformation of the generated photons using a convex lens of focal length, $f$=50 mm in $f$-$f$ optical system configuration, is recorded using EMCCD Each pixel represents a particular transverse position $r$ of the down converted photons, which corresponds to a transverse momentum value, $K =[\omega_{dc}/(c f )]r$, where, $\omega_{dc}$ is the frequency of the down converted photons, $c$ is the velocity of light in free space and $f$ is the focal length of the Fourier transforming lens. In the current study we have used two topological charges, $l=0$ (Gaussian beam) and $l=3$ and varied their ring radius $\rho_r$ for different axicon positions (D-value) and recorded the angular spectrum of the down converted photons with the results shown in Fig. 5. First row, (a)-(c), and second row, (d)-(f), of Fig. 5 represents the angular spectrum of the down converted photons corresponding to pump radius, $\rho_r$=0.3, 0.75 and 1.18 mm for $l=0$ and $l=3$ respectively. $K_x$ and $K_y$ are the transverse momentum components of the down converted photons. As evident from the Fig. 5, for a given annular ring radius $\rho_r$ of both the topological charges, $l=0$ and $l=3$, the down converted photons have identical transverse profiles with intense inner and outer rings and intensity gradient in between. Therefore, it can be concluded that the OAM of the input pump beam has no effect in the transverse distribution of the down converted photons. However, it should be noted that the sum of the OAMs of signal and idler photons of each pair will be equal to the OAM of the input pump photon[23]. The dark spot at the centre of each transverse momentum distribution can be attributed to the back ground correction used to avoid the pump beam leaked through the interference filter. Although the increase in pump beam radius does not modify the size of the transverse momentum profile of the down converted photons, at lower pump beam size the down converted photons have asymmetric transverse momentum distribution. Such asymmetry arising from the Poynting vector walk-off in the birefringent crystals are observed for tightly focused beams of radius smaller or equivalent to the lateral displacement of the beam, $L \tan\rho_o$ in the nonlinear crystal of length, $L$ and birefringence parameter, $\rho_o$. For $L$=5 mm and $\rho_o$=67.44 mrad of the BBO crystal, the $L \tan\rho_o$ value is calculated to be 0.34 mm. As a result we can expect asymmetric transverse distribution in down converted photons for pump beam radius, $\rho_r$=0.3 mm.

However, the asymmetry in the transverse momentum distribution of the down converted photons is detrimental for entangled photon sources[23], and for all practical purposes, one need symmetric distribution. Therefore, we have studied the asymmetry in transverse momentum distribution of the down converted photons quantitatively with the radius of the perfect vortex beam. The results are shown in Fig. 6. The asymmetry factor, $\xi$ is defined as $\xi$=1-$a/b$, where $a$ and $b$ representing the separation between the inner and out radii on left and right sides of the angular spectrum of the down converted photons respectively as shown in inset of the Fig. 6. We have calculated the asymmetry factor, $\xi$, for different size of the perfect vortex of order $l$=3. As evident from the Fig. 6, the $\xi$ value decreases linearly with the increase of perfect vortex ring radius. Although it is expected to have symmetric angular spectrum of the down converted photons for the input vortex ring radius $\rho_r > L \tan\rho_o$ (>0.34 mm), we still observe a certain degree of asymmetry in angular spectrum at higher values of $\rho_r$. However, for $\rho_r$ =1.18 mm (~3 x $L \tan\rho_o$) we have observed symmetric ($\xi$<10%) angular spectrum of the down converted photons. Such study ascertain the need of proper beam diameter of the input structured beam (here, perfect vortex beam) depending upon the birefringence and length of the nonlinear crystal for entangled photon sources.

**Methods**

The schematic of the experimental setup is shown in Fig. 1 (a). A pump diode UV laser (Toptica- TopMode 405) of 100 mW at 405nm with spectral width ~12 MHz is used as the pump laser. The power to the nonlinear crystal is controlled by using a combination of half-wave plate ($\lambda$/2) and polarising beam splitter (PBS1) cube. Two spiral phase plates, SPP1 and SPP2 are used to convert Gaussian beam into optical vortices (LG beams) of order, $l$=1 and $l$=2, respectively, at conversion efficiency >95%. Due to limited availability of SPPs producing vortices of order $l$>2, we have used a vortex-doubler setup[21] comprised of a polarizing beam splitter cube (PBS 2), quarter-wave plate ($\lambda$/4), and dielectric coated mirror, M1, with high reflectivity at 405nm. The working principle of the vortex doubler[21] can be understood as follows. The Gaussian beam in horizontally polarization passes through the PBS2. In forward pass through the SPP1 the Gaussian beam acquires spiral phase corresponding to vorticity of order $l$ =+1 (say). The $\lambda$/4 plate having its axes at 45° to its polarization axis converts the polarization of the beam into left circular polarization.

The mirror, M, reverses both the polarization handedness (left circular to right circular) and sign of the vorticity ($l$ =+1 to -1) of the reflected beam. The right circularly polarized beam on return pass through the $\lambda/4$ plate gets converted into vertical polarization. To the return beam ($l$ = -1), the direction of phase variation of the SPP1 is opposite to that of the forward pass. Therefore, the return beam acquires additional spiral phase resulting total vorticity, $l$ = -2, twice the phase winding of the SPP placed inside. Since the return beam have vertical polarization, it gets reflected by the PBS2. Using different combinations of SPP1 and SPP2, placed inside and outside of the vortex-doubler we have generated optical vortices of orders (topological charges) up to $l$=6. An antireflection coated 1" diameter axicon made of BK-7 glass having apex angle 178° is used to convert the vortex beam of order $l$ into BG beam of the same order. As shown in Fig. 1 (b), the Fourier transforming lens (*f=200 mm*) placed before the axicon with a separation of *f-D* (*D* is the distance of the axicon from Fourier plane) produces the perfect vortices (Fig. 1c) at its back focal plane. To measure the change in the vortex radius we have place the CCD camera (SP620U, Ophir) at the back focal plane of the Fourier lens and varied the position of the axicon. A 5 mm long and 4 x 5 mm aperture, $BaB_2O_4$ (BBO) is used as the nonlinear crystal for down conversion of the pump beam (Fig. 1d). The crystal is cut at $\theta$=29.9°, internal angle with the normal incidence, for type-I ($e \rightarrow o+o$) non-collinear phase-matching to produce degenerate signal and idler photons at 810 nm at an angle 3° with respect the pump beam. Due to the circular symmetry, the photons are generated in a cone of apex angle 6° and the photons of each pair are situated in diametrically opposite points. A second lens, $f$=50 mm Fourier transforms the generated SPDC ring (Fig. 1e) which is recorded using electron multiplying CCD (EMCCD, Andor-iXon Ultra 897) after filtering the pump photon using an interference filter, IF of 10 nm bandwidth centred at 810 nm. The EMCCD has 512×512 pixels with pixel size of 16 micron. To suppress background noise the EMCCD was operated at -80°C. We have taken images by accumulating 20 frames exposure time of 0.5s. The $\lambda/2$ placed after the mirror M2 is to adjust the polarization of the input beam depending on the crystal orientation for perfect phase matching.

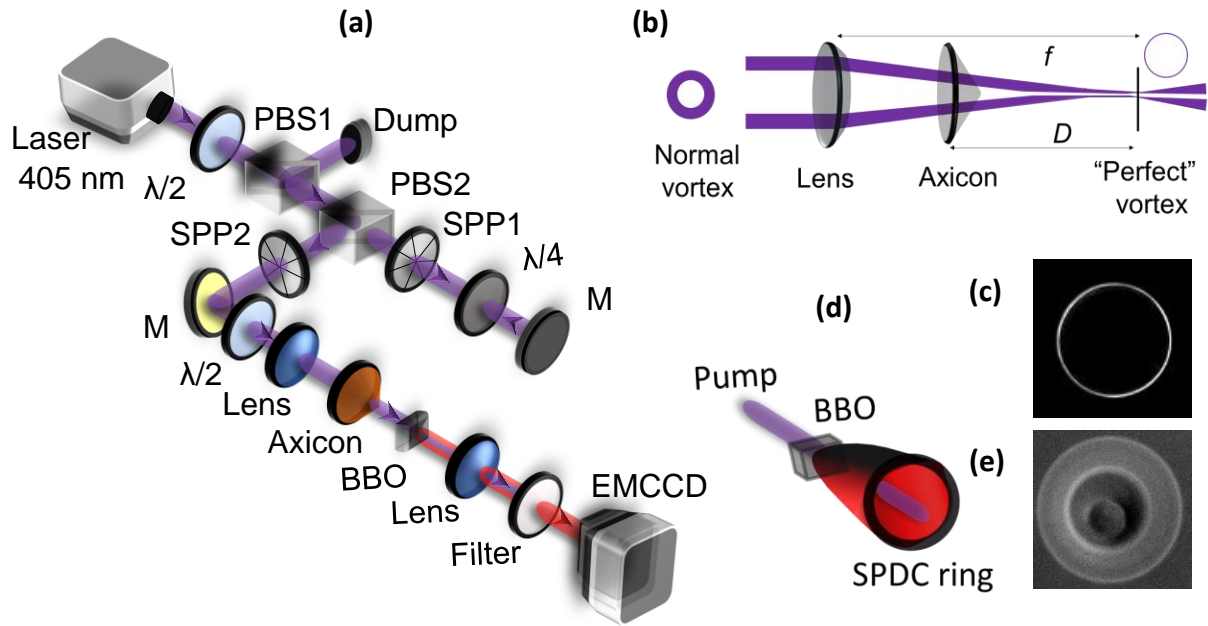

Fig. 1. Experimental layout for variable size perfect vortex and SPDC as described in detail in the methods section. (a) Schematic of the experimental setup. λ/2, half wave plate; PBS1-2, polarizing beam splitter cube; SPP1-2, spiral phase plate; λ/4, quarter wave plate; M1-2, Mirrors; Axicon, to generate BG beam; BBO, nonlinear crystal for down conversion; Filter, interference filter; EMCCD, electron multiplying CCD. (b) Pictorial representation of the generation of variable size perfect vortex by using lens-axicon combination. (c) Experimentally measured intensity profile of perfect vortex. (d) Schematic of generation of SPDC ring. (e) Experimentally measured angular spectrum of the down converted photons while pumping with perfect vortex.

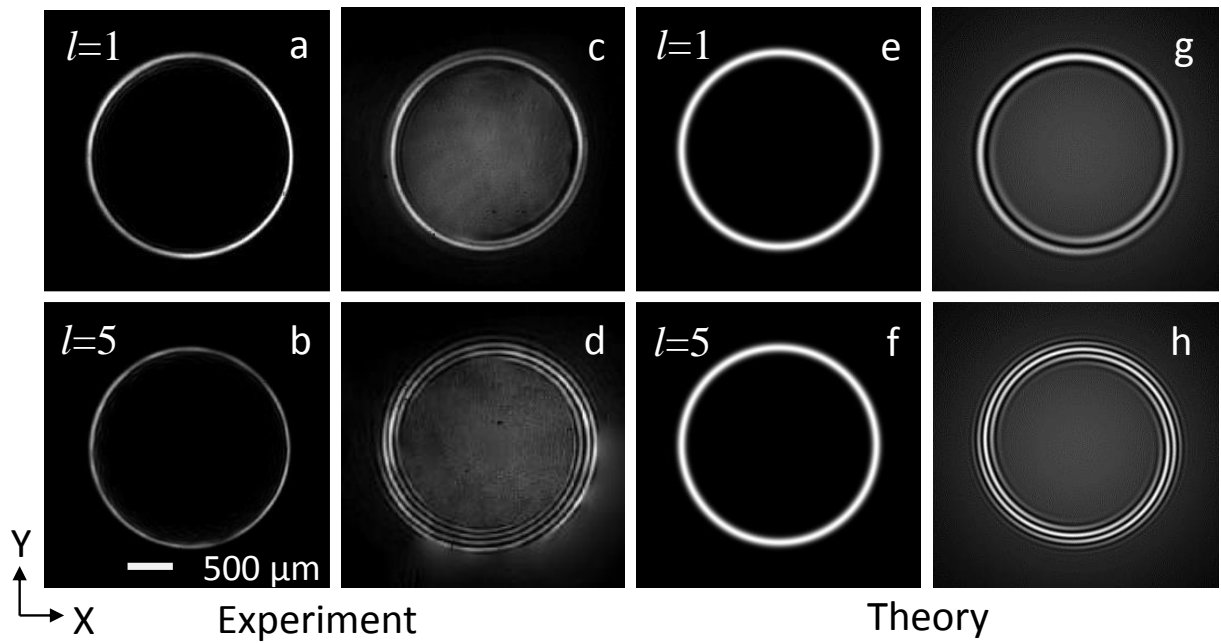

Fig. 2. Intensity distribution of the perfect vortices and their interference pattern with Gaussian beam. Experimental intensity distribution (a), (b) and corresponding interference pattern (c),(d) of the perfect vortex of order $l$=1 and $l$=5. Theoretical intensity distribution (e),(f) and interference pattern (g),(h) of the perfect vortex of order $l$=1 and $l$=5.

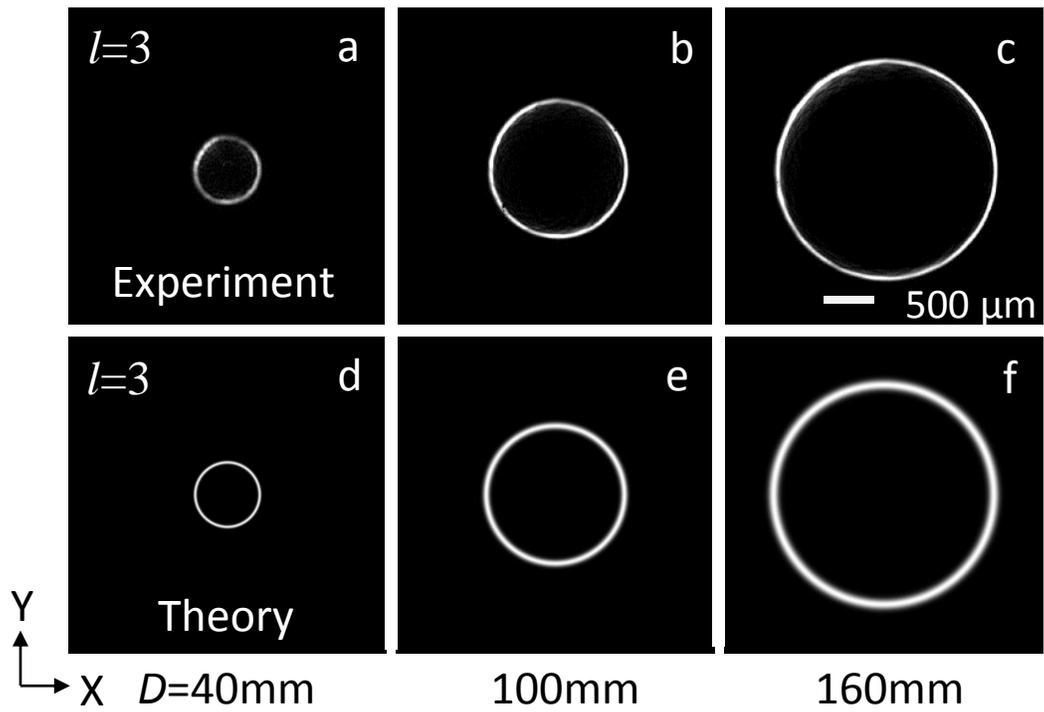

Fig. 3. Experimental and theoretical images of the perfect vortex of order *l*=3 for different position of the axicon. First row and second row respectively represent the experimental and theoretical images of the perfect vortex for *D* value (a,d) 40mm, (b,e) 100mm, (c,f) 160 mm.

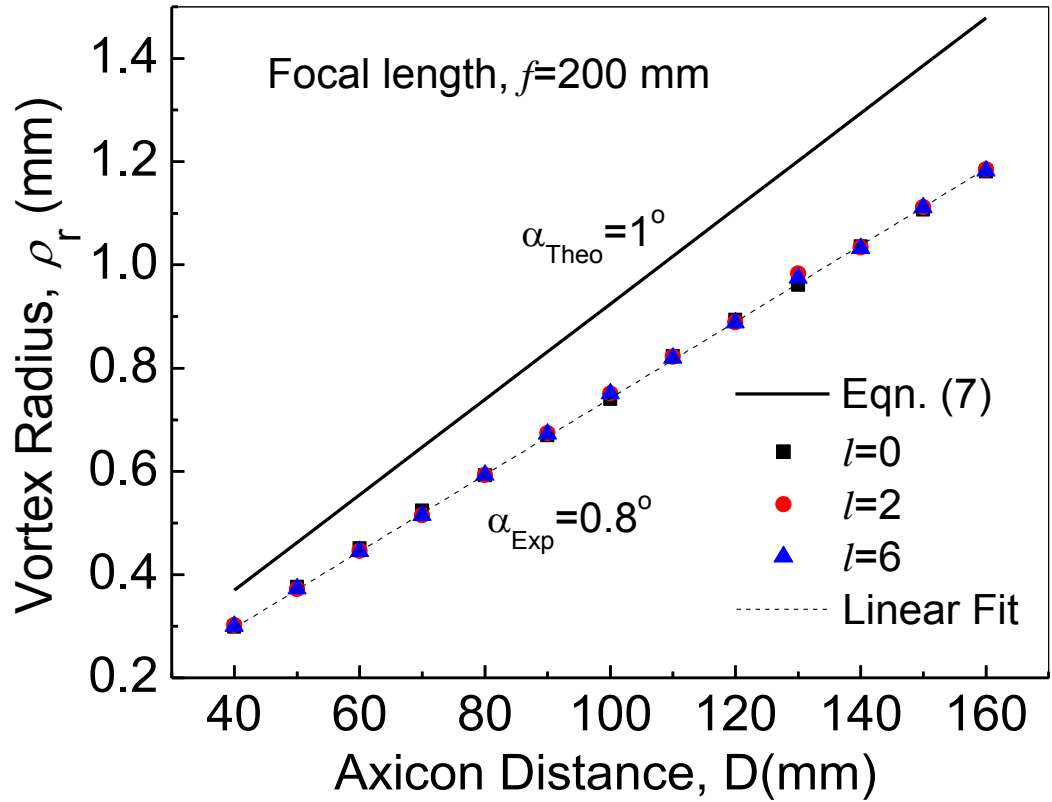

Fig. 4. Variation of perfect vortex radius with distance *D* for vortex order 0, 2 and 6. The solid points and dotted line represent the experimental data and linear fit respectively. Solid line represents theoretical beam radius of the perfect vortex calculated using Eq. (7) and axicon parameters.

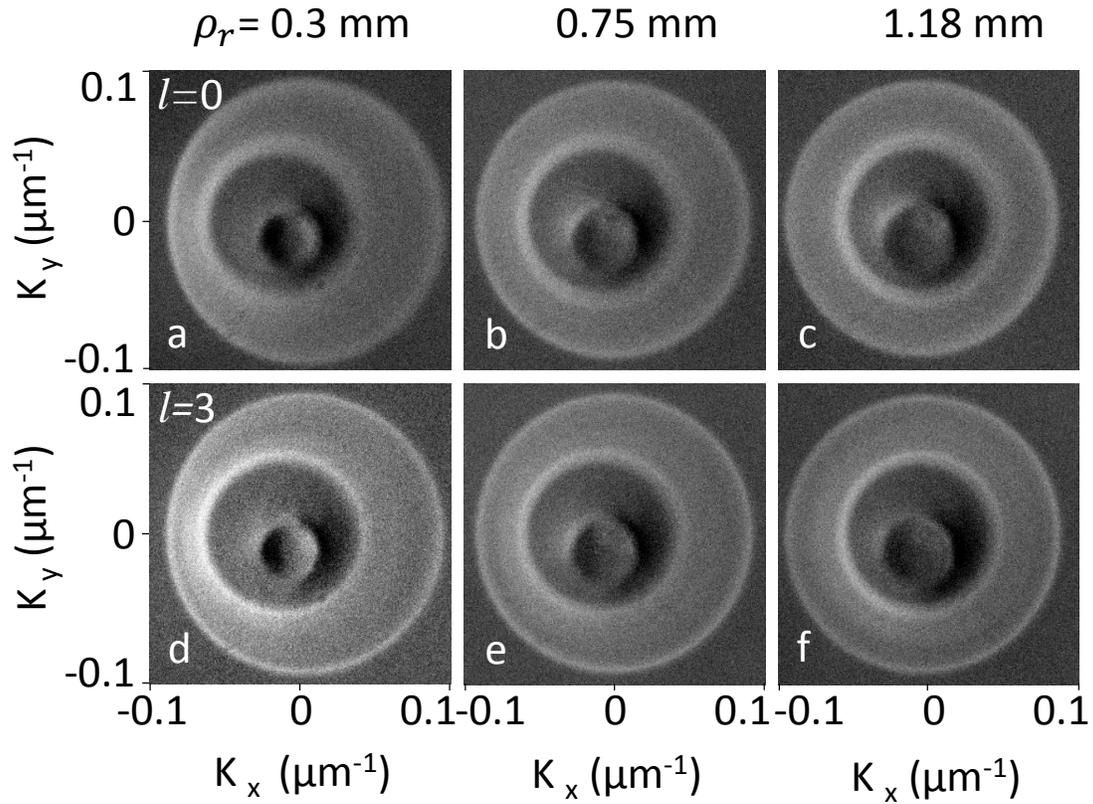

Fig. 5. Angular spectrum (transverse momentum) of the down converted photons recorded using EMCCD for different pump beam radius and vortex orders. First row and second row correspond to the angular spectrum of the down converted photons for pump orders $l=0$ and $l=3$ respectively. First, second and third column represent transverse momentum distribution of the down converted photons for pump beam radius 0.3, 0.75, and 1.18 mm respectively. $K_x$ and $K_y$ are the transverse momentum components.

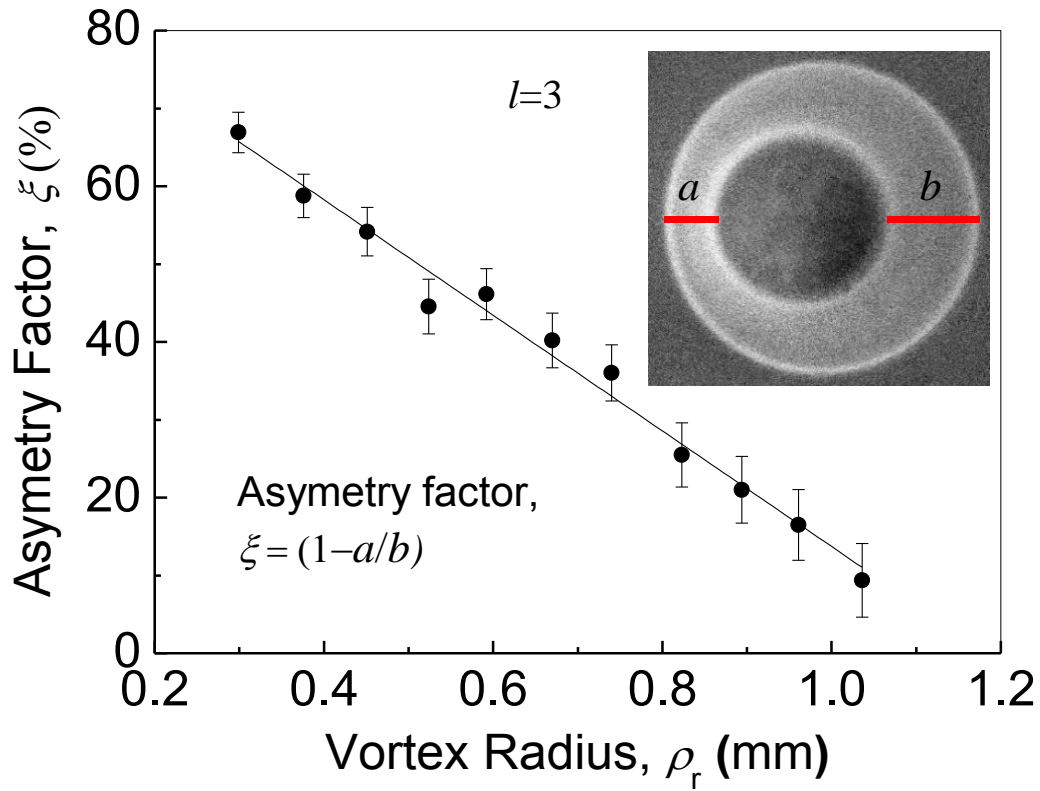

Fig. 6. Dependence of asymmetry of the angular spectrum of the down converted photons on the radius of pump vortex beam. Asymmetry parameter is defined as $\xi = (1-a/b)$, where, '$a$' and '$b$' are quantified as the separation between two rings on left and right side of the transverse momentum distribution of the down converted photons (inset).